# Eshelby's method for unidirectional periodic composites

Guo-Qing Gu [1], En-Bo Wei [2,a]

[1] *School of Computer Science and Technology, East China Normal University, Shanghai 20062, People's Republic of China*

[2] *Institute of Oceanology, Chinese Academy of Sciences and Qingdao National Laboratory for Marine Science and Technology, Qingdao 266071, People's Republic of China*

[a] Author to whom correspondence should be addressed: ebwei@qdio.ac.cn

**Abstract** Open boundary conditions are always used in investigating the effective properties of composites. In this paper, Eshelby's transformation field method is developed to deal with the effective response of unidirectional periodic composites having an open boundary. In the method, Hermite polynomials are used to cope with the open boundary conditions of the perturbation fields induced by the inclusions. The transformation fields are introduced in the composite system to meet the interface conditions between complex structure inclusions and matrix. As an example, Eshelby's method is used to estimate the effective responses of two-dimensional unidirectional periodic dielectric composites having an open boundary. The validity is verified by comparing the effective responses calculated by the method with the exact solutions of dilute limit. It is shown that the method is valid to solve the open boundary problem of unidirectional periodic composites having complex geometric inclusions.

**Keywords**: Transformation field, Open boundary. Effective dielectric property, Unidirectional periodic composites

## 1. Introduction

For general composites, there are many theoretical methods for solving various problems of composites [1-3]. Owing to the complex interactions of complex structure inclusions and matrix, the field distributions and the macroscopic response are difficult to be exactly derived. Eshelby's method [4,5] proposed for calculating the effective elastic tensors of the ellipsoidal composite is extensively studied for various problem of composites[6-13].

In Eshelby's method, the transformation constant strain inside the ellipsoid was introduced to set up the stress equations in the ellipsoidal inclusion region on basis of the continuous surface traction across the interfaces of ellipsoidal inclusions. Generally, the concept of transformation strain tensor, which is no longer a constant tensor, can be extended to the non-ellipsoidal elastic composite [14]. For example, Nemat-Nasser et al [15,16]. had developed Eshelby's method to estimate the effective module of an elastic periodic composite using Fourier series of field quantities [17]. Subsequently, Gu et al [18, 19]. first extended the method to estimate the effective electric conductivity of periodic composites, the effective viscosity of periodic suspensions [20], the dispersion relation of photonic band structure of periodic dielectrics and the effective piezoelectric properties of piezoelectric periodic composites [21,22]. Moreover, Zhou [23] calculated the effective elastic responses of periodic composites having multiple inhomogeneous inclusions of arbitrary shapes. For the graded composites, Eshelby's method was used to obtain the effective dielectric constants of periodic composites having the inclusions of arbitrary gradient and geometry [24,25]. However, although Eshelby's transformation field method has successfully solved the above periodic boundary problem, the open boundary problem is more difficult to be solved. The open boundary is the primitive problem of predicting the effective properties of the composites even though Rayleigh proposed the periodic model of the composites [26]. Applied an external field to a composite, the open boundary conditions are always used since the perturbation fields, induced by inclusions, tend to zero at infinite point, which is far from the inclusion region. The fields in an infinite point regarded as the external applied field is an intrinsic open boundary condition in physics. Therefore, it is expected to develop Eshelby's method to solve the open boundary problem of composites having arbitrary inclusion shapes [27-33].

In this paper, the transformation field method of the complex inclusion geometric composites having an open boundary is proposed. Without loss of generality, the effective dielectric response of the two-dimensional unidirectional periodic isotropic dielectric composites having an open boundary is estimated. In Sec.2, the transformation field method is derived for the open boundary problem of unidirectional periodic isotropic dielectric composites having arbitrary inclusion shapes. To meet the open boundary conditions, Hermite polynomials are used to express the transformation fields and the perturbation fields. Moreover, a set of algebraic equations is given for solving the transformation field. In Sec.3, the effective dielectric formulas are derived on the basis of the transformation fields, which can be solved in Sec. 2. In Sec.4, the validity of our method is verified by comparing the effective dielectric responses calculated by the method with the exact solution of dilute limit. Finally, a brief conclusion is given in Sec.5.

## 2. Formulation

Consider an isotropic dielectric composite consisting of periodically distributed long fiber inclusions of common geometry along one direction, where the unit cell has an open boundary along the other directions, and contains one or several long fibers, as an example of a long cylindrical composite with unidirectional period along the $y$-direction is shown in Fig.1, where the periodic length of the unit cell along the $y$-direction is $2L$. Here note that the cross section of a long fiber is a close surface of arbitrary inclusion geometry, and the coordinate origin is located at the gravity center of the cross section of an inclusion.

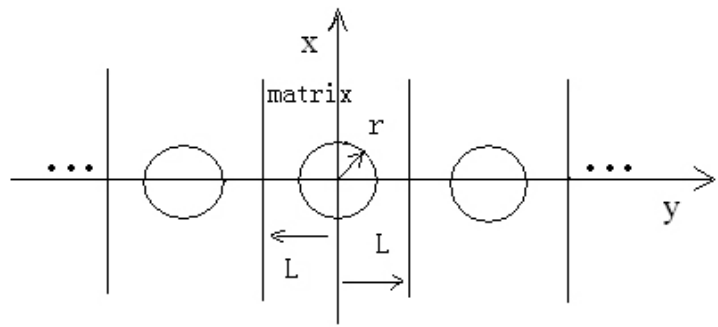

**Fig.1**. Cross section schematic diagram of an infinite long cylindrical composite having the y-directional period structure.

For the two-dimensional problem in Cartesian coordinates $(x, y)$, we assume that the isotropic constitutive relations of both matrix and fiber inclusions are linear,

$$D_k^\alpha = \varepsilon^\alpha E_k(x,y), \tag{1}$$

where $k = x, y$. $\alpha = h, i$ denote the quantities of the matrix and inclusions, respectively. $D$, $E$ and $\varepsilon$ are the electrical displacement, electrical field and dielectric constant, respectively. The governing electrostatic equations are $\nabla \cdot D = 0$ and $\nabla \times E = 0$. If an external uniform electric field $E_y^0$ is applied to the composites along the $y$-direction, the perturbation electric field $E_k^p$, induced by the inclusions, will occur in whole composite region. In the inclusion region $\Omega_i$ and the host region $\Omega_h$ of the unit cell, the constitutive relations are given by $D_k^i = \varepsilon^i (E_k^0 + E_k^p)$ and $D_k^h = \varepsilon^h (E_k^0 + E_k^p)$, respectively. To avoid matching the interfaces continuous conditions of the electrical displacements between inclusion and matrix, a transformation electrical field $E_k^*$ is introduced to the composites according to Eshelby's method such that its components obey the following equations in the matrix and inclusion regions, respectively [4,15,18],

$$E_k^*(x,y) = 0 \quad \text{in} \quad \Omega_h, \tag{2}$$

$$\varepsilon^i [E_k^0 + E_k^p(x,y)] = \varepsilon^h [E_k^0 + E_k^p(x,y) - E_k^*(x,y)] \quad \text{in} \quad \Omega_i, \tag{3}$$

where $k = x, y$. Based on Eqs.(2) and (3), a unified constitutive relation is obtained in whole composite region $\Omega = \Omega_i + \Omega_h$.

$$D_k = \varepsilon^h [E_k^0 + E_k^p(x,y) - E_k^*(x,y)] \quad \text{in} \quad \Omega. \tag{4}$$

Then, the governing equation is rewritten as follows

$$\nabla_k \{\varepsilon^h [E_k^0 + E_k^p(x,y) - E_k^*(x,y)]\} = 0. \tag{5}$$

Clearly, Eq.(5) indicates the relationships between the perturbation electric field and the transformation field.

Since the composite is periodic only along the $y$-direction, the perturbation potential $\Phi(x,y)$ (i.e. the perturbation electric field, $\vec{E}^p(x,y) = -\nabla \Phi$) and the transformation electric field $E_k^*(x,y)$ are periodic in the $y$-direction and the open boundary in the $x$

-direction. In order to express the perturbation potential by the transformation field, the perturbation potential $\Phi(x,y)$ and the transformation field $E_k^*(x,y)$ are expanded in terms of Hermite polynomials about the direction variable $x$ of the open boundary and Fourier series about the periodic directional variable $y$.

$$\Phi(x,y) = \sum_{m=0}^{\infty} \sum_{n=-\infty}^{\infty} \Phi_{m,n} u_m(x) v_n(y), \tag{6}$$

$$E_x^*(x,y) = \sum_{m=0}^{\infty} \sum_{n=-\infty}^{\infty} \alpha_{m,n} u_m(x) v_n(y), \tag{7}$$

$$E_y^*(x,y) = \sum_{m=0}^{\infty} \sum_{n=-\infty}^{\infty} \beta_{m,n} u_m(x) v_n(y), \tag{8}$$

where $\Phi_{m,n}$, $\alpha_{m,n}$ and $\beta_{m,n}$ are unknown constants. The orthogonal function $u_m(x)$ is $u_m(x) = N_m H_m(x)\exp(-\frac{1}{2}x^2)$, where $H_m(x)$ is Hermite polynomial and the coefficient $N_m$ is $N_m = (\frac{1}{2^m m!\sqrt{\pi}})^{1/2}$. The orthogonal function $v_n(y)$ is $v_n(y) = \frac{1}{\sqrt{2L}}\exp(i\xi_n y)$ with $\xi_n = \frac{2n\pi}{2L}$. Here note that $x \in (-\infty,\infty)$ and $y \in [-L,L], L>0$. Substituting Eqs.(6), (7) and (8) into Eq.(5), and considering the orthogonality of function $v_n(y)$, we have

$$\sum_{m=0}^{\infty}[\Phi_{m,n}\frac{d^2 u_m(x)}{dx^2} - \xi_n^2 \Phi_{m,n} u_m(x)] = \sum_{m=0}^{\infty}[-\alpha_{m,n}\frac{du_m(x)}{dx} - i\xi_n \beta_{m,n} u_m(x)]. \tag{9}$$

Moreover, the following Eqs.(10) and (11) are derived by Hermite polynomial,

$$\frac{du_m(x)}{dx} = \tilde{N}_m^+ u_{m+1}(x) + \tilde{N}_m^- u_{m-1}(x), \tag{10}$$

$$\frac{d^2 u_m(x)}{dx^2} = \tilde{M}_m^+ u_{m+2}(x) + \tilde{M}_m^0 u_m(x) + \tilde{M}_m^- u_{m-2}(x), \tag{11}$$

where $\tilde{N}_m^+ = -\frac{1}{2}N_m/N_{m+1}$, $\tilde{N}_m^- = mN_m/N_{m+1}$, $\tilde{M}_m^+ = \tilde{N}_m^+ \tilde{N}_{m+1}^+$, $\tilde{M}_m^0 = \tilde{N}_m^+ \tilde{N}_{m+1}^- + \tilde{N}_m^- \tilde{N}_{m-1}^+$ and $\tilde{M}_m^- = \tilde{N}_m^- \tilde{N}_{m-1}^-$. Substituting Eqs.(10) and (11) into Eq.(9), we obtain Eq.(12) with respect to the function $u_m(x)$,

$$\sum_{m=0}^{\infty}[\Phi_{m,n}\widetilde{M}_m^+ u_{m+2}(x)+\Phi_{m,n}(\widetilde{M}_m^0-\xi_n^2)u_m(x)+\Phi_{m,n}\widetilde{M}_m^- u_{m-2}(x)]$$
$$=\sum_{m=0}^{\infty}[-\alpha_{m,n}\widetilde{N}_m^+ u_{m+1}(x)-\alpha_{m,n}\widetilde{N}_m^- u_{m-1}-i\xi_n\beta_{m,n}u_m(x)]$$
(12)

where $n=0,\pm 1,\pm 2,\cdots,\pm N$. Clearly, if the infinite series of Eq.(12) is approximated to the $M$-th order of the perturbation potential coefficient $\Phi_{M,n}$, we can get a set of equations about the coefficients $\Phi_{m,n}$, $\alpha_{m,n}$ and $\beta_{m,n}$ by comparing the coefficients of orthogonal function $u_m(x)$ ($m=0,1,2,\cdots,M$) in both sides of Eq.(12). Then, the coefficient equations are given by

$$\Phi_{0,n}(\widetilde{M}_0^0-\xi_n^2)+\Phi_{2,n}\widetilde{M}_2^- = -(\alpha_{1,n}\widetilde{N}_1^- + i\xi_n\beta_{0,n})$$

$$\Phi_{1,n}(\widetilde{M}_1^0-\xi_n^2)+\Phi_{3,n}\widetilde{M}_3^- = -(\alpha_{0,n}\widetilde{N}_0^+ + \alpha_{2,n}\widetilde{N}_2^- + i\xi_n\beta_{1,n})$$

$$\Phi_{0,n}\widetilde{M}_0^+ +\Phi_{2,n}(\widetilde{M}_2^0-\xi_n^2)+\Phi_{4,n}\widetilde{M}_4^- = -(\alpha_{1,n}\widetilde{N}_1^+ + \alpha_{3,n}\widetilde{N}_3^- + i\xi_n\beta_{2,n})$$

$$\Phi_{1,n}\widetilde{M}_1^+ +\Phi_{3,n}(\widetilde{M}_3^0-\xi_n^2)+\Phi_{5,n}\widetilde{M}_5^- = -(\alpha_{2,n}\widetilde{N}_2^+ + \alpha_{4,n}\widetilde{N}_4^- + i\xi_n\beta_{3,n})$$

$$\cdots$$

$$\Phi_{M-2,n}\widetilde{M}_{M-2}^+ +\Phi_{M,n}(\widetilde{M}_M^0-\xi_n^2)+\Phi_{M+2,n}\widetilde{M}_{M+2}^- = -(\alpha_{M-1,n}\widetilde{N}_{M-1}^+ + \alpha_{M+1,n}\widetilde{N}_{M+1}^- + i\xi_n\beta_{M,n}). \quad (13)$$

To solve the coefficients $\Phi_{m,n}$, the closed equations can be determined by truncating the $m$ th-order of the coefficient $\Phi_{m,n}$ in Eq.(13). Generally, the solutions of coefficients $\Phi_{m,n}$ can be written as the following form,

$$\Phi_{m,n}=w_{m,n}\sum_{l=0}^{K}(\gamma_{m,l,n}^{\alpha}\alpha_{l,n}+\gamma_{m,l,n}^{\beta}\beta_{l,n}), \quad (14)$$

where $K$ is the approximation order of the function $u_m(x)$. The coefficients $w_{m,n}$, $\gamma_{m,l,n}^{\alpha}$ and $\gamma_{m,l,n}^{\beta}$ can be derived from Eq.(13).

For example, approximating the coefficient $\Phi_{m,n}$ to its second order (i.e. $\Phi_{m,n}=0$ for $m>2$), we have the following equations from the first three equations of Eq.(13),

$$\Phi_{0,n}(\widetilde{M}_0^0-\xi_n^2)+\Phi_{2,n}\widetilde{M}_2^- = -(\alpha_{1,n}\widetilde{N}_1^- + i\xi_n\beta_{0,n})$$

$$\Phi_{1,n}(\widetilde{M}_1^0 - \xi_n^2) = -(\alpha_{0,n}\widetilde{N}_0^+ + \alpha_{2,n}\widetilde{N}_2^- + i\xi_n\beta_{1,n})$$

$$\Phi_{0,n}\widetilde{M}_0^+ + \Phi_{2,n}(\widetilde{M}_2^0 - \xi_n^2) = -(\alpha_{1,n}\widetilde{N}_1^+ + \alpha_{3,n}\widetilde{N}_3^- + i\xi_n\beta_{2,n}). \tag{15}$$

The approximation solutions of the second order $\Phi_{2,n}$ is derived by Eq.(15) with the linear relations of the coefficients $\alpha_{m,n}$ and $\beta_{m,n}$. These approximation solutions are finally given by,

$$\Phi_{0,n} = w_{0,n}(\gamma_{0,1,n}^\alpha \alpha_{1,n} + \gamma_{0,3,n}^\alpha \alpha_{3,n} + \gamma_{0,0,n}^\beta \beta_{0,n} + \gamma_{0,2,n}^\beta \beta_{2,n}),$$

$$\Phi_{1,n} = w_{1,n}(\gamma_{1,0,n}^\alpha \alpha_{0,n} + \gamma_{1,2,n}^\alpha \alpha_{2,n} + \gamma_{1,1,n}^\beta \beta_{1,n}),$$

$$\Phi_{2,n} = w_{2,n}(\gamma_{2,1,n}^\alpha \alpha_{1,n} + \gamma_{2,3,n}^\alpha \alpha_{3,n} + \gamma_{2,0,n}^\beta \beta_{0,n} + \gamma_{2,2,n}^\beta \beta_{2,n}), \tag{16}$$

where $w_{0,n} = [\widetilde{M}_0^+ \widetilde{M}_2^- - (\widetilde{M}_0^0 - \xi_n^2)(\widetilde{M}_2^0 - \xi_n^2)]^{-1}$, $\gamma_{0,1,n}^\alpha = \widetilde{N}_1^- \widetilde{M}_2^0 - \xi_n^2 \widetilde{N}_1^- - \widetilde{M}_2^- \widetilde{N}_1^+$,

$\gamma_{0,3,n}^\alpha = -\widetilde{M}_2^- \widetilde{N}_3^-$, $\gamma_{0,0,n}^\beta = i\xi_n(\widetilde{M}_2^0 - \xi_n^2)$, $\gamma_{0,2,n}^\beta = i\xi_n \widetilde{M}_2^-$, $w_{1,n} = (\widetilde{M}_0^0 - \xi_n^2)^{-1}$, $\gamma_{1,0,n}^\alpha = -\widetilde{N}_0^+$,

$\gamma_{1,2,n}^\alpha = -\widetilde{N}_2^-$, $\gamma_{1,1,n}^\beta = -i\xi_n$, $w_{2,n} = 1/\widetilde{M}_2^-$, $s_{0,n} = (\widetilde{M}_0^0 - \xi_n^2)$, $\gamma_{2,1,n}^\alpha = (s_{0,n}w_{0,n}\gamma_{0,1,n}^\alpha + \widetilde{N}_1^-)$,

$\gamma_{2,3,n}^\alpha = s_{0,n}w_{0,n}\gamma_{0,3,n}$, $\gamma_{2,0,n}^\beta = (s_{0,n}w_{0,n}\gamma_{0,0,n} + i\xi_n)$, $\gamma_{2,2,n}^\beta = s_{0,n}w_{0,n}\gamma_{0,2,n}$.

Next, the transformation electric field $E_k^*(x,y)$ can be determined by Eq.(3). Substituting Eqs.(14), (6) and (10) into Eq.(3), we obtain the following equations in the inclusion region $\Omega_i$,

$$\varepsilon^h E_x^*(x,y) - (\varepsilon^h - \varepsilon^i)E_x^0 = \\ (\varepsilon^h - \varepsilon^i)\sum_{n=-\infty}^{\infty}\sum_{m=0}^{\infty}\{[w_{m,n}\sum_{l=0}^{p}(-\gamma_{m,l,n}^\alpha \alpha_{l,n} - \gamma_{m,l,n}^\beta \beta_{l,n})]D_{m,n}(x,y)\} \tag{17}$$

$$\varepsilon^h E_y^*(x,y) - (\varepsilon^h - \varepsilon^i)E_y^0 = \\ (\varepsilon^h - \varepsilon^i)\sum_{n=-\infty}^{\infty}\sum_{m=0}^{\infty}\{[w_{m,n}\sum_{l=0}^{p}(-\gamma_{m,l,n}^\alpha \alpha_{l,n} - \gamma_{m,l,n}^\beta \beta_{l,n})]u_m(x)(i\xi_n)v_n(y)\} \tag{18}$$

where $D_{m,n}(x,y) = [\widetilde{N}_m^+ u_{m+1}(x) + \widetilde{N}_m^- u_{m-1}(x)]v_n(y)$. $\alpha_{l,n} = \int_{\Omega_i} u_l(x)v_{-n}(y)E_x^*(x,y)dxdy$ and $\beta_{l,n} = \int_{\Omega_i} u_l(x)v_{-n}(y)E_y^*(x,y)dxdy$, which are derived from Eqs.(7) and (8), respectively.

To solve the transformation fields, we expand the transformation field as a series, such as a power series,

$$E_k^*(x, y) = \sum_{i,j=0}^{\infty} C_k^{ij} \left(\frac{x}{L}\right)^i \left(\frac{y}{L}\right)^j, \tag{19}$$

where $k = x, y$. $C_k^{ij}$ is an unknown coefficient. Substituting Eq.(19) into Eqs. (17) and (18), and multiplying both sides of the resulting equations by a term $\left(\frac{x}{L}\right)^k \left(\frac{y}{L}\right)^l$, and then integrating these equations over the inclusion region $\Omega_i$, we get a set of linear algebra equations about the unknown coefficient $C_k^{ij}$,

$$(\varepsilon^h - \varepsilon^i)\left[\sum_{i,j=0}^{\infty}\sum_{s=0}^{p} C_x^{ij} H_\alpha^{ij,kl}(s) + \sum_{i,j=0}^{\infty}\sum_{s=0}^{p} C_y^{ij} H_\beta^{ij,kl}(s)\right] + \varepsilon^h \sum_{i,j=0}^{\infty} C_x^{ij} A^{k+i, j+l} = (\varepsilon^h - \varepsilon^i) E_x^0 A^{k,l}, \tag{20}$$

$$(\varepsilon^h - \varepsilon^i)\left[\sum_{i,j=0}^{\infty}\sum_{s=0}^{p} C_x^{ij} F_\alpha^{ij,kl}(s) + \sum_{i,j=0}^{\infty}\sum_{s=0}^{p} C_y^{ij} F_\beta^{ij,kl}(s)\right] + \varepsilon^h \sum_{i,j=0}^{\infty} C_y^{ij} A^{k+i, j+l} = (\varepsilon^h - \varepsilon^i) E_y^0 A^{k,l}, \tag{21}$$

where. $H_\alpha^{ij,kl}(s) = \sum_{m=0, n=-\infty}^{\infty} \gamma_{m,s,n}^{\alpha} w_{m,n} D_{m,n}^{kl} G^{ij}(s, -n)$,

$H_\beta^{ij,kl}(s) = \sum_{m=0, n=-\infty}^{\infty} \gamma_{m,s,n}^{\beta} w_{m,n} D_{m,n}^{kl} G^{ij}(s, -n)$,

$F_\alpha^{ij,kl}(s) = \sum_{m=0, n=-\infty}^{\infty} \gamma_{m,s,n}^{\alpha} w_{m,n} i\xi_n G^{ij}(s, -n) G^{kl}(m, n)$,

$F_\beta^{ij,kl}(s) = \sum_{m=0, n=-\infty}^{\infty} \gamma_{m,s,n}^{\beta} w_{m,n} i\xi_n G^{ij}(s, -n) G^{kl}(m, n)$,

$G^{ij}(m, n) = \int_{\Omega_i} (x/L)^i (y/L)^j u_m(x) v_n(y) dxdy$,

$A^{i,j} = \int_{\Omega_i} (x/L)^i (y/L)^j dxdy$,

$D_{m,n}^{kl} = \tilde{N}_m^+ G^{kl}(m+1, n) + \tilde{N}_m^- G^{kl}(m-1, n)$.

Here note that, in above derivation, the following formulas are used,

$$\alpha_{l,n} = \int_{\Omega_i} u_l(x) v_{-n}(y) E_x^*(x, y) dxdy = \sum_{i,j=0}^{\infty} C_x^{ij} G^{ij}(l, -n),$$

$$\beta_{l,n} = \int_{\Omega_i} u_l(x)v_{-n}(y)E_y^*(x,y)dxdy = \sum_{i,j=0}^{\infty} C_y^{ij} G^{ij}(l,-n).$$

Thus, the unknown coefficients $C_k^{ij}$ can be solved with the closed Eqs.(20) and (21) if the number of Eqs.(20) and (21), multiplied by the term $(\frac{x}{L})^k (\frac{y}{L})^l$ ($k,l = 0,1,2,\ldots,S$ and $S$ notes the order of power-law for the closure equations), equals to the number of the unknown coefficient $C_k^{ij}$. It is clear that there is not any limitation for the inclusion geometric shapes. Thus, our formulas of Eqs.(20) and (21) can be used to the composites having the inclusions of arbitrary shapes due to the complex boundary differential problem of composites translated to the integration problem of inclusion regions.

## 3. Effective dielectric responses

Based on the electric fields of composites region and the constitutive relations, the effective dielectric constant $\varepsilon_{ij}^e$ of two-dimensional composites can be defined as the following form,

$$\overline{D}_k = \varepsilon_{kx}^e \overline{E}_x + \varepsilon_{ky}^e \overline{E}_y, \tag{22}$$

where subscript $k = x, y$. $\overline{A}$ notes the spatial average of quantity $A$ over the whole composite region $\Omega = \Omega_h + \Omega_i$. For isotropic composites, we have the following Eq.(23) for the volume average of electric displacement components,

$$\frac{1}{V}\int_{\Omega}(D_k - \varepsilon^h E_k)dV = \frac{1}{V}\int_{\Omega_i}(D_k^i - \varepsilon^h E_k)dV = \overline{D}_k - \varepsilon^h \overline{E}_k, \tag{23}$$

where $\Omega_i$ is the regions of inclusions and $V$ is the volume of whole composite region. Substituting Eq.(22) into Eq.(23), we have,

$$\frac{1}{V}\int_{\Omega_i}(D_k^i - \varepsilon^h E_k)dV = \varepsilon_{kx}^e \overline{E}_x + \varepsilon_{ky}^e \overline{E}_y - \varepsilon^h \overline{E}_k. \tag{24}$$

Then, for the $y$-direction external electrical field $E_y^0$, we can regard the external electrical field as the average field. Taking $\overline{E}_y = E_y^0$ and $\overline{E}_x = 0$ in Eq.(24) for $k = y$, we obtain the effective dielectric constant formula $\varepsilon_{yy}^e$,

$$\varepsilon_{yy}^e = \varepsilon^h + \frac{f}{V_i E_y^0} \int_{\Omega_i} (\varepsilon^i E_y - \varepsilon^h E_y) dV, \qquad (25)$$

where $V_i$ is the volume of inclusions. $f = V_i / V$ is the volume fraction of inclusions in the whole composite regions. Furthermore, taking $\overline{E}_y = E_y^0$ and $\overline{E}_x = 0$ in Eq.(24) for $k = x$, we have the effective dielectric constant formula $\varepsilon_{xy}^e$,

$$\varepsilon_{xy}^e = \frac{f}{V_i E_y^0} \int_{\Omega_i} (\varepsilon^i E_x - \varepsilon^h E_x) dV, \qquad (26)$$

Moreover, based on the Eq.(3), the electrical fields $E_k^i(x, y)$ of inclusion regions are given by the transformation electrical field,

$$E_k^i(x, y) = E_k^0 + E_k^p(x, y) = \frac{\varepsilon^h}{\varepsilon^h - \varepsilon^i} E_k^*(x, y), \qquad (27)$$

Finally, Substituting Eq.(27) into Eqs.(25) and (26), we can estimate the effective responses of the composite on the basis of the transformation field solutions given in above Section.

## 4. Numerical discussion

To verify the validity of our method, as an example, the effective dielectric responses of a long cylindrical unidirectional periodic composite having an open boundary along the $x$-direction (see Fig.1) is estimated and compared with the analytical results of the cylindrical composites in the dilute limit [34]. In the numerical calculation, the second order approximation solutions of the perturbation potential $\Phi_{2,n}$, given in Eq.(16), are used under an external uniform electric field $E_y^0$ along the $y$-direction. The dielectric constants of both isotropic matrix and inclusion are $\varepsilon^h = 20$ and $\varepsilon^i = 3n$, respectively, where $n$ is the factor of inclusion dielectric constants. The radius $r$ of the cylinder inclusion is $r = 2$, and the semi-periodic length $L$ of the unit cell in the $y$-direction is $L = kr$ ($k > 1$).

In Fig.2, the effective dielectric constant ratios $\varepsilon_{yy}^e / \varepsilon^h$ changing with the dielectric factor $n$ of inclusions for different semi-periodic length $L$ are shown, where the volume faction of dilute inclusion is $f = 0.05$. It is clear that the effective dielectric constants asymptotically converge to the exact results ("dilute" in Fig.2 notes the exact dilute solutions) of cylindrical composites in the dilute limit with the length $L$ increasing [34]. Good agreements are obtained comparing our results with those exact results for the length $L$ larger than $13r$. The results indicate that the interactions between inclusions become very weak with the increase of the parameter $L$. However, in the case of the smaller dielectric contrast ratios (i.e. $\varepsilon^i/\varepsilon^h$ less than a critical value), the effects of the periodic length $L$ on the effective response is negligible by comparison with the case of the larger contrast ratios. It implies that the coupling interactions of the material properties and the composite structure are also important in predicting the effective responses, especially for the larger dielectric contrast ratios $\varepsilon^i/\varepsilon^h$. In addition, it is noted that the effective dielectric $\varepsilon_{xy}^e$ is zero in our numerical calculation because of the isotropic material and symmetrical structure. Hence, the numerical results demonstrate that the extended Eshelby's method is valid to cope with the periodic composites having an open boundary.

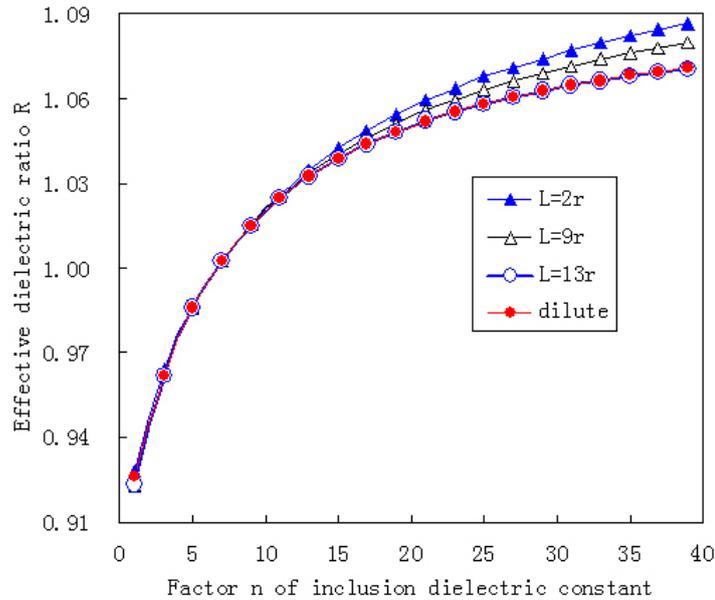

**Fig.2**. Effective dielectric ratio $R$ (i.e. $\varepsilon_{yy}^e/\varepsilon^h$) versus the factor $n$ of inclusion dielectric constant, where the radius of the inclusion cylindrical

cross section is $r=2$ and the length of the y-direction periodic unit cell is $2L$. The isotropic dielectric constants of matrix and cylindrical materials are $\varepsilon^h = 20$ and $\varepsilon^i = 3n$, respectively. The volume fraction $f$ of inclusions is 0.05. In this figure, "dilute" notes the exact results of cylindrical composites in the dilute limit.

## 5. Conclusions

For unidirectional periodic composites having an open boundary, Eshelby's transformation field method is developed to investigate its effective response. In the method, the transformation field is introduced in the composite system to deal with the complex interface conditions between the inclusion and the matrix. The open boundary conditions are treated by Hermite polynomials. The transformation electrical field is expanded by Fourier series in the periodic direction. As an example, the method is applied to the isotropic cylindrical periodic composite having an open boundary along $y$-direction. The effective dielectric responses are approximately estimated by truncating the series of Hermite polynomials. The numerical results show that the developed Eshelby's method is very useful to cope with the open boundary problem of unidirectional periodic composites. The effects of the coupling interactions between the material properties and the composite structure cannot be neglected on the effective properties and the perturbation field. Finally, as a brief conclusion, it is noted that the developed method is valid for the kind of the unidirectional periodic composites having various inclusion shapes since the method is not any limitations for the inclusion shapes.

**Acknowledgements**

This work was supported by NSFC (Grant No. 42276178).

## References

[1] G. W. Milton, *The theory of composites* (Cambridge University Press, Cambridge,

2002).

[2] R. Hill, J. Mech. Phys. Solids **13** (1965) 213.

[3] M. Landstorfer, B. Prifling, V. Schmidt, J. Comput. Phys. **431** (2021) 11007.

[4] J. D. Eshelby, Proc. R. Soc. London, Ser. A **241** (1957) 376.

[5] J. D. Eshelby, Proc. R. Soc. London, Ser. A **252** (1959) 561.

[6] J. W. Ju, T. M. Chen, Acta Mechanica **103** (1994) 103.

[7] X. L. Gao, M. Q. Liu, J. Mech. Phys. Solids **60** (2012) 261.

[8] M. J. Huang, W. N. Zou, Q. S. Zheng, Int. J. Eng. Sci. **47** (2009) 1240.

[9] J. F. Yu, X. H. Ni, Z. G. Cheng, W. H. He, X. Q. Liu, Y. W. Fu, Strength of Materials **53** (2021) 342.

[10] J. D. Eshelby, Philos. Mag. **6** (1961) 953.

[11] Q. S. Zheng, Z. H. Zhao, D. X. Du, J. Mech. Phys. Solids **54** (2006) 368.

[12] T. Y. Yuan, K. F. Huang, J. X. Wang, J. Mech. Phys. Solids, **158** (2022) 104648.

[13] W. N. Zou, Q. C. He, Q. S. zheng, Proc. R. Soc. A **469** (2013) 20130221

[14] S. Trotta, G. Zuccaro, S. Sessa, F. Marmo, L. Rosati, Compos. Part B-Eng. **144** (2018) 267.

[15] S. Nemat-Nasser, M. Taya, Q. Appl. Math. **39** (1981) 43.

[16] S. Nemat-Nasser, T. Iwakuma, M. Hejazi, Mech. Mater. **1** (1982) 239.

[17] T. Iwakuma, S. Nemat-Nasser, Comput. Struct.**16** (1983)13.

[18] G. Q. Gu, R. Tao, Phys. Rev. B **37** (1988) 8612.

[19] G. Q. Gu, R. Tao, Acta Phys. Sin. **37** (1988) 582.

[20] G. Q. Gu, R. Tao, Sci. China, Ser. A: Math., Phys., Astron. Technol. Sci. **32**  (1989) 1186.

[21] G. Q. Gu, P. M. Hui, C. Xu, W. C. Woo, Solid State Commun. **120** (2001) 483.

[22] E. B. Wei, Y. M. Poon, F. G, Shin, G. Q. Gu, Phys. Rev. B **74** (2006) 014107.

[23] K. Zhou, Acta Mech. **223** (2012) 293.

[24] E. B. Wei, G. Q. Gu, K. W. Yu, Phys. Rev. B **76** (2007) 134206.

[25] E. B. Wei, G. Q. Gu, Y. M. Poon, Phys. Rev. B **77** (2008) 104204.

[26] L. Rayleigh, Philos. Mag. **34** (1892) 481.

[27] S. Yang, J. Wang, G. L. Dai, F. B. Yang, J. P. Huang, Phys. Rep. **908** (2021) 1.

[28] D. J. Bergman and K. J. Dunn, Phys. Rev. B **45** (1992)13262.

[29] D. J. Bergman and D. Stroud, Solid State Phys. **46** (1992) 147.

[30] J. P. Huang and K. W. Yu, Phys. Rep. **431** (2006) 87.

[31] A. Mejdoubi and C. Brosseau, Phys. Rev. E **73** (2006) 031405.

[32] P. A. Martin, J. Eng. Math. **42**,133 (2002).

[33] B. Jiang, D. N. Fang, K. C. Hwang, Int. J. Solid Struct. **36** (1999) 2707.

[34] E. B. Wei, J. B. Song, G. Q. Gu, J. Appl. Phys. **95** (2004) 1377.